\documentclass[a4paper,11pt]{article}
\usepackage{pos}
\usepackage{float}
\usepackage[rflt]{floatflt}

\usepackage{seqsplit}

\newcommand{\ep}{\varepsilon}
\newcommand{\Li}{{\rm Li}}
\newcommand{\HA}{{\rm H}}

\title{
Holonomic techniques for massive 3-loop form factors: the gluonic case\footnote{DESY-26-108, RISC Report number 26-13, PoS(LL2026)071
}}
\ShortTitle{Holonomic techniques for massive 3-loop form factors: the gluonic case}

\author[a,b]{J.~Bl\"umlein}
\author[c]{A.~De Freitas}
\author[a]{P.~Marquard}
\author[c]{J.~Obrovsky}
\author*[c]{C.~Schneider}

\affiliation[a]{Deutsches Elektronen-Synchrotron DESY, Platanenallee 6, 15738 Zeuthen, Germany}

\affiliation[b]{Institut f\"ur Theoretische Physik III, IV, TU Dortmund, Otto-Hahn
Stra\ss{}e 4, 44227 Dortmund, Germany}

\affiliation[c]{Johannes Kepler University Linz, Research Institute for Symbolic
	Computation (RISC), Altenberger Stra\ss{}e 69, A-4040, Linz, Austria}

\emailAdd{Johannes.Bluemlein@desy.de}
\emailAdd{Peter.Marquard@desy.de}
\emailAdd{afreitas@risc.jku.at}
\emailAdd{Jakob.Obrovsky@risc.jku.at}
\emailAdd{Carsten.Schneider@risc.jku.at}

\abstract{Massive three-loop form factors for vector, axial-vector, scalar, and pseudoscalar currents are vital for precision collider phenomenology. Extending the quarkonic framework to the more complex gluonic case, in Ref.~\cite{BFMOS:26} we computed these contributions widely using advanced computer algebra, automated guessing, and large-scale PSLQ searches. Here, we outline the core strategy of the large-moment method and illustrate its main challenges, including solving record-sized recurrences and differential equations. Our high-precision evaluation provides an exact analytic representation around $s=0$ in terms of MZVs. We achieved a complete analytic series expansion around $s=\pm\infty$ for the first time, which depends on MZVs and three constants out of
higher number spaces, along with reliable analytic continuations across $s \in (-\infty, \infty)$, serving as a landmark benchmark for modern symbolic computation.}

\FullConference{Loops and Legs in Quantum Field Theory (LL2026)\\
12-17, April, 2026\\
Bayreuth, Germany\\}

\begin{document}
\maketitle


\section{Introduction}

Massive form factors for vector, axial-vector, scalar, and pseudoscalar currents parameterize the interaction between virtual bosons and massive quark pairs in both Quantum Chromodynamics (QCD) and Quantum Electrodynamics (QED). Acting as indispensable components in heavy-quark precision calculations, these form factors feed into numerous observables vital to collider phenomenology, such as (virtual) Higgs decays and heavy-flavor production. Matching the high accuracy of modern experimental measurements demands matching theoretical precision, driving the evaluation of these quantities to higher loop orders.

Computing three-loop massive form factors represents a particularly challenging problem due to the large number of contributing Feynman diagrams and the complexity of the underlying Feynman integrals, which involve structures beyond standard harmonic polylogarithms. While complete semi--analytic results~\cite{Fael:2022miw,Egner:2022jot,Fael:2022rgm,Schonwald:2022djs,Fael:2023zqr,Schonwald:2023uel},
and partial analytic results are available in the literature~\cite{Gluza:2009yy,Henn:2016kjz,Henn:2016tyf,Ahmed:2017gyt,Ablinger:2018yae,Ablinger:2018zwz,Lee:2018nxa,Lee:2018rgs,Blumlein:2018tmz,Blumlein:2019oas}, we provided a complete analytic result in the high-energy limit for the quarkonic case in~\cite{Blumlein:2023uuq} and the gluonic case in~\cite{BFMOS:26}. 
The calculation of the gluonic contributions follows the framework previously established for the quarkonic case. However, because the gluonic case proved to be significantly more complex, its evaluation required new technological advancements and more sophisticated computer algebra techniques. In this note, we summarize the principal steps of our calculation using the method of large moments~\cite{Blumlein:2017dxp} and illustrate the technical challenges with concrete examples.

\section{Calculation steps of the holonomic approach and challenges}
\label{sec:1}

\begin{center}
	\begin{figure}[h!]
		\begin{center}
			\begin{minipage}[c]{0.15\linewidth}
				\includegraphics[width=1\textwidth]{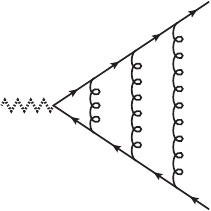}
				\vspace*{-10mm}
				\begin{center}
					{\footnotesize{(a)}}
				\end{center}
			\end{minipage}
			\hspace*{0.01\linewidth}
			\begin{minipage}[c]{0.15\linewidth}
				\includegraphics[width=1\textwidth]{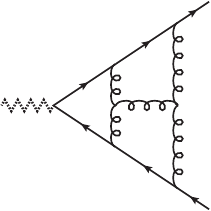}
				\vspace*{-10mm}
				\begin{center}
					{\footnotesize (b)}
				\end{center}
			\end{minipage}
			\hspace*{0.01\linewidth}
			\begin{minipage}[c]{0.15\linewidth}
				\includegraphics[width=1\textwidth]{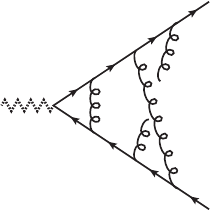}
				\vspace*{-10mm}
				\begin{center}
					{\footnotesize (c)}
				\end{center}
			\end{minipage}
			\hspace*{0.01\linewidth}
			\begin{minipage}[c]{0.15\linewidth}
				\includegraphics[width=1\textwidth]{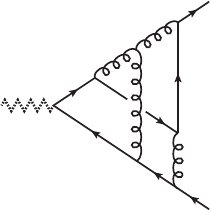}
				\vspace*{-10mm}
				\begin{center}
					{\footnotesize (d)}
				\end{center}
			\end{minipage}
			\hspace*{0.01\linewidth}
			\begin{minipage}[c]{0.15\linewidth}
				\includegraphics[width=1\textwidth]{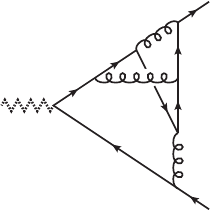}
				\vspace*{-10mm}
				\begin{center}
					{\footnotesize (e)}
				\end{center}
			\end{minipage}
		\end{center}
		\caption{\small Sample of diagrams required for the calculation of three-loop massive form factors. The solid arrow lines represent massive quarks and the curly lines the gluons,
			while the dotted zigzag line represents any of the possible external currents. 
		}
		\label{fig:samplediagrams}
	\end{figure}
\end{center}

The gluonic contributions to the form factors arise exclusively from diagrams in which only gluons are exchanged between the heavy quarks; for typical examples see Fig.~\ref{fig:samplediagrams}.

These diagrams are expressed in terms of dimensionally regularized three-loop Feynman integrals in $D=4-2\ep$ space-time dimensions.
The form factors depend on the dimensionless variable $s$, defined as
\begin{equation}
	\label{eq:kin}
	s = \frac{q^2}{m^2},
\end{equation}
where $q^2$ denotes the virtuality of the incoming current and $m$ the heavy-quark mass. Alternatively, we employ the dimensionless variable $x$, related to $s$ through
\begin{equation}
	\label{eq:x2s}
	s = -\frac{(1-x)^2}{x} \implies x = \frac{\sqrt{4-s}-\sqrt{-s}}{\sqrt{4-s}+\sqrt{-s}};
\end{equation}
for further details we refer to~\cite{BFMOS:26}.

Using integration-by-parts (IBP) reduction techniques~\cite{Chetyrkin:1981qh,Laporta:2001dd}, in particular the software package {\tt Crusher}~\cite{CRUSHER},
the form factors $F_I$ are expressed as 
a linear combination of 417 master integrals $M_i$ (MIs):
\begin{equation}\label{Equ:LinComMi}
	F_{I}(s)=c_1(\ep,s)M_1(\ep,s)+\dots+c_{417}(\ep,s)M_{417}(\ep,s)
\end{equation}
with rational functions $c_i(\ep,s)$ in $s$ and $\ep$. An important byproduct is that the $M_i$ themselves can be described by a coupled system of first-order linear differential equations. In addition, the initial values for the differential equations are specified at $s=0$ ($x=1$), with corresponding expansions given as power series in either $s$ or $y=1-x$. In short, the full problem is uniquely determined by~\eqref{Equ:LinComMi}, the coupled systems in the $M_i$ and their initial values.

Using this input, the overall calculation strategy is based on the large moment method~\cite{Blumlein:2023uuq} enhanced by advanced analytic continuation. It can be summarized as follows~\cite{BFMOS:26}.

\begin{enumerate}
	\item[\textbf{1}] \textbf{Generate Master Integral Series:} Compute $\rho$ coefficients in $s$ (or in $1-x$) up to the required order in $\ep$ for the power series representations of the master integrals by solving the coupled system of first-order linear differential equations obtained from IBP reductions.
	
	\item[\textbf{2}] \textbf{Construct Form Factor Series:} Substitute the truncated power series solutions of the master integrals into Eq.~\eqref{Equ:LinComMi} and expand to obtain $\rho$ coefficients in $s$ (or in $1-x$) for the form factors. The resulting coefficients are linear combinations of constants 
	\begin{equation}
		\left\{1, \, \zeta_2, \,  \zeta_3, \, \ln(2) \zeta_2, \, \zeta_2^2, \, \Li_4\left(\frac{1}{2}\right), \, \ln^4(2), \, \ln^2(2) \zeta_2, \, \zeta_2 \zeta_3, \, \zeta_5 \right\}
		\label{eq:MZV-list}
	\end{equation}
	and color factors; $\zeta_z=\zeta(z)=\sum_{i=1}^{\infty}\frac{1}{i^z}$ denotes the Riemann zeta-function and $\Li_n(x) = \sum_{k=1}^{\infty} \frac{x^k}{k^n}$ the classical polylogarithm.
	
	\item[\textbf{3}] \textbf{Guess Recurrences and Differential Equations:} For each finite list of rational numbers $(F(n))_{n=0}^{\rho-1}$, guess a linear recurrence relation and a linear differential equation satisfied by the power series $f(s)=\sum_{n=0}^{\infty} F(n) s^n$ (or $f(x) = \sum_{n=0}^{\infty} F(n) (1-x)^n$).
	
	\item[\textbf{4}] \textbf{Solve Factorizable Systems:} For components where the recurrence/differential equations factorize into first-order operators, solve the recurrences in terms of iterative sums, and convert its power series representation to iterative integrals.
	
	\item[\textbf{5}] \textbf{Analytic Continuation:} For non-first-order factorizable contributions, employ Frobenius' method and extensions, see~\cite{FROBENIUS,COHI,vdH:singhol,mezzarobba2016,AKR:22} and references therein, to construct high-order local series expansions around regular and singular points ($s \to \pm\infty$, $s=4$, $s=16$), matching overlapping expansions to achieve analytic continuation across the full kinematic range $s \in (-\infty, \infty)$.
\end{enumerate}

For smaller values $\rho$, say $\rho=1000$ it was straightforward to execute Steps~1 and~2 to obtain $1000$ coefficients for all of the form factors in $x$ or in $s$ using the techniques described in~\cite{RefinedLM1:20,Blumlein:2023uuq}.
In particular, for the contributions up to $O(\ep^{-1})$, we could carry out Step~3 to guess all the underlying recurrences and to solve them in closed form in Step~4. In this way, we rediscovered the results in terms of generalized harmonic sums~\cite{Vermaseren:1998uu,Blumlein:1998if,Moch:2001zr,Ablinger:2011te,Ablinger:2013cf} and harmonic polylogarithms~\cite{Remiddi:1999ew}
that have been calculated earlier in~\cite{Gluza:2009yy,Henn:2016kjz,Henn:2016tyf,Ahmed:2017gyt,Ablinger:2018yae,Ablinger:2018zwz,Lee:2018nxa,Lee:2018rgs,Blumlein:2018tmz,Blumlein:2019oas}. In particular, we succeeded in computing for $O(\ep^0)$ the coefficients of the color factors involving $\zeta_5$ and $\zeta_2\,\zeta_3$. For instance, consider the form factor
\begin{equation}
f(x)=\sum_{k=0}^{\infty}F(n)(1-x)^n
\end{equation}
for the color factor $C_F^3\zeta_5$ at the projection
$g_1$ of the axial-vector form factor; see~\cite{BFMOS:26}.
Using the first 30 values of the coefficients
\small
\begin{align}
(F(n))_{n\geq0}=&\Big(-9760,0,-\frac{6560}{3},-\frac{6560}{3},-2184,-\frac{6544}{3},-\frac{45848}{21},-\frac{15328}{7},-\frac{19796}{9},-\frac{19904}{9},\nonumber\\
&-\frac{513928}{231},-\frac{172376}{77},-\frac{2899232}{1287},-\frac{2268976}{1001},-\frac{20545076}{9009},-\frac{984128}{429},-\frac{20785178}{9009},\nonumber\\
&-\frac{20900336}{9009},-\frac{616991548}{264537},-\frac{757990580}{323323},-\frac{6855857452}{2909907},-\frac{6888767972}{2909907},-\frac{159175786100}{66927861},\nonumber\\
&-\frac{159888055660}{66927861},-\frac{42257750536}{17612595},-\frac{62019593516}{25741485},-\frac{2428552034444}{1003917915},-\frac{106002999364}{43648605},\nonumber\\
&-\frac{23657503150756}{9704539845},-\frac{23744630290016}{9704539845},\dots\Big)
\end{align}
\normalsize
produced by Steps~1 and~2 we employed the \texttt{ore\_algebra} package~\cite{SageOre,GSAGE} developed in Sage and guessed the recurrence
\begin{multline}
-(n-1) \left(2 n^4+n^3+5 n^2-60 n+48\right) F(n)\\
+2 (4 n^5-n^4-n^3-107 n^2+69 n+156)F(n+1)\\
+(-10 n^5-n^4+34 n^3+187 n^2+342 n-1272) F(n+2)\\
+2 (n+4)(2 n^4-7 n^3+14 n^2-75 n+114)F(n+3)=0;
\end{multline}
note that the other 970 values (which have not been used for guessing) confirm that the found result is reliable. Next, we used the summation package \texttt{Sigma}~\cite{SIG1,SIG2} in the setting of
difference rings~\cite{KARR,BRONSTEIN,ABRAMOV,CS4,CS5,CS8,CS12} to solve the found recurrence yielding the closed form expression
\begin{equation}
F(n)=-\frac{160 \big(
	13-22 n+6 n^2+7 n^3\big)}{(-1+n) n (1+n)}
-320 S_1({n})
-25 2^{4-n} S_1({{2},n})
\end{equation}
in terms of the harmonic numbers
$S_1(n)=\sum_{k=1}^n\frac1k$ and the generalized harmonic sum $S_1({{2},n})=\sum_{k=1}^n\frac{2^k}{k}$. Finally, using the package \texttt{HarmonicSums}~\cite{Ablinger:2010kw,Ablinger:2013hcp} and the technologies given in~\cite{Vermaseren:1998uu,Blumlein:1998if,Moch:2001zr,Ablinger:2011te,Ablinger:2013cf,Remiddi:1999ew} we derive the representation
\begin{equation}
f(x)=-\frac{160 \big(
	7+30 x+7 x^2\big)}{x}
-\frac{160 \big(
	2+11 x+8 x^2+11 x^3+2 x^4\big) \ln(x)}{(-1+x) x (1+x)}.
\end{equation}
Note that the number of rational coefficients needed to guess the recurrences
depends on the constants~\eqref{eq:MZV-list}
under consideration. 

For the solvable parts, we used the expansion in $1-x$; compare the example above. As it turns out, 
in order to guess the linear recurrences, one needs fewer coefficients of the expansion in $s$ than in $1-x$. For further considerations we decided therefore to focus on the expansions in $s$.
For the constants $\zeta_2^2$, $\ln^2(2) \zeta_2$, $\ln^4(2)$, $\Li_4\left(\frac{1}{2}\right)$, $\zeta_2 \zeta_3$ and $\zeta_5$,
500 rational coefficients suffice. For the case $\ln(2) \zeta_2$ we needed around 6000 coefficients, and for
$\zeta_2$ and $\zeta_3$ we required 10000 and 12000 coefficients, respectively. In the most complicated case, namely the constant contribution $1$,
we could get the recursions using 26000 coefficients. For the underlying techniques to derive so many coefficients in Steps~1 and~2 we refer to~\cite{BFMOS:26} and the used ideas given in~\cite{RefinedLM1:20,Blumlein:2023uuq,RefinedLM2:24}. 

In the following we want to give some more details on the most complicated problem: the color factor $C_F^3$ 
at the projection
$g_1$ of the axial-vector form factor. Here we needed around $25300$ values\footnote{In a first step we calculated only 25000 values and failed to get the recurrence for this particular instance. By new technologies described in~\cite{BFMOS:26} we produced afterwards 40000 values (which was clearly overshooting) to guarantee the final success.} to guess the linear recurrence. 
To illustrate the size of these rational numbers in this case, the 26000th value printed in tiny font size requires 18 pages (163879 decimals digits for the numerator and 156054 for the denominator)\footnote{The 40000th coefficient (that we computed from the IBP relations) in printed form requires 40 pages (251779 digits for the numerator and 239755 digits for the denominator.}.
For the guessing of the linear recurrence we used an enhanced version of the \texttt{ore\_algebra} package. The found recurrence (in printed form it needs $\sim$650000 A4-pages) is of the form
\begin{equation}
a_0(n) F(n)+a_1(n)F(n+1)+\dots+a_{91}(n)F(n+91)=0
\end{equation}
where the coefficients $a_i(n)$ are polynomials with integer coefficients with degree up to $4078$. For instance, the constant term of the polynomial $a_{91}(n)\in\mathbb Z[n]$ is given in Fig.~\ref{fig:coeff} which requires $8207$ decimal digits.

\begin{center}
	\begin{figure}[t]
		\begin{center}
\begin{minipage}[t]{14cm}\tiny
		\seqsplit{-774728490714239632605056533471825979785387145167981470593826940582796\
			5750752933574420096562229991294477862717073198152676091023269272320063\
			0485661823726556356204008752949666798017642632839185287749454251519651\
			6833578790273493791767050920379554835251825175805998197160190831642835\
			5857748115086379919054727150637432800362594260334478395203480919173263\
			2140146340463394547661510621426446813254825336124680804998849567073367\
			6899489160468744411549754738472405332054635355487054057616565930696285\
			1226879803003389936972437742066072673074031692970060479125720040241854\
			8991861745601040637960000135873438491407975567775207268965185724007240\
			8962891714733881424433362429354656718937677414503126676340840921521210\
			7581673780955372163435238535660463727646215990397932252822432712715100\
			9297573788670812199977273740311509071909886140074498835819469571569780\
			2619036803558925507929283307224421642379508274956072525028505674639256\
			1745539565934366475507301977072572414715268869577342960801110477247749\
			3568072229185549197313890493781948335106988508966883431259292198395613\
			8051188433210497675541583262059638268586211644701118875235220703574962\
			2269004684192657953593660900375797356705998257000681409608560387707949\
			1167112205921599271513253323399945264809210397014195768871502006945386\
			0791279545421097102762497013317495309638652633129717395799228644263000\
			6581792387116046651415609402159176172825775505585244791279825367755512\
			3544859301155329676090177835863585301593367101263547462325527405138092\
			2097409383492441098607864500395622101942569341324077867402527207422609\
			1232749047029906167614680148458839122055056519617735258691608162881837\
			5652460645635611784314030965864425231794516266141691653063689426866027\
			1174561638569611568617490169706291855693144618509940153528519993975269\
			6465383731291360819821573149575544061298319808439909115763103775974165\
			8494566494142686244141176281746064301994095097310963520027222967664264\
			6834134572235007684447754871583492932276891950773179746296400103889155\
			2888909063940296046269603942899578229840898704896228238015779262437328\
			5143297154853772670519152055148935585088925390591315353842998764682732\
			1026257419451918921689668595582509133337950389991262905953186939760353\
			6452587833943450035602638709923238149400583533695882337232620434957613\
			3273523656700185378833358125423529333192767360954588046186784615293038\
			4349566387926853372012419239419921059467947900576064967497257826499040\
			3472219809366790147463958605751739172139043925538753283600710247821728\
			8245916588539182451109509741037216747422138068883719046241026383768318\
			4041282893107176046729559207452543531790874672565220584132292808858374\
			8625675745509959932617065255530202637061159082150066036395045180622179\
			0943216926967185957328068465887798362784366644770562037091081637414667\
			1192639701959199990301448688922622794757755573270151680850553209835330\
			7875699226272825494893376374255702090977483431204926921810171780275899\
			5276437163646217761318220508584853658308433743828702291331751499629379\
			5728434832552243835359455277720221269605606810258698652243331848410020\
			9699837351216775638632358519112366759702331139044817959661161767469801\
			8135155998167173665354908924919054138338895971892015406108193643150793\
			5047832005032613678745734636538393816730372700816437716540285600434742\
			9536577405713665703683877316733740924638641731175812014203261449477727\
			8338550436310492487184400863772244333409709540881034962356888905328319\
			7852758369761950929108794332575166463892665386027370774096565848565944\
			6209931572730892744237835035939619265952807421207794826136747213668067\
			4555429066458744956594762316376940508096388805517471781817362720928871\
			7797032325947203654480645284020261261616614593819921088807803577082235\
			8132678090172221936929558036850652091173738442643784949787594609335679\
			3495626848868478168288667048476732171279963769528730368528071336358025\
			0000923702864889179870015604148768450116020907095368831373629504680649\
			8071301198150497338222979483171381840488298535276612101026961114847428\
			3316648911012247322051400893722929583194231324130581556727213900010055\
			8758990554450974790910231547331542982980412131411762673217726128618390\
			2095723207290500861344845879315119474679825474120823583532123145751521\
			4655352696304687613335977756494082530712502600559353763729990877213478\
			7676808642767723952008969807756328666444238643328302225075820969204945\
			2126816532474579921359754199407087412777781987797198473695782875749088\
			0030164248455334905475129437030281929065541362149030870651995588726982\
			7520043652170450462382419121557073421404109752033369157195109778116449\
			6748142546660145771714831524511626474098065456977376200768668078656731\
			0586376237255252987348202331074070174838282116537555555249410584765612\
			1872451754128157994465672391578154193436267498613747351544623923996857\
			1675566360606148521711390336399078385775613545652959625759522990665087\
			2050834746960880405130032113412131638323524301467297457388133892515772\
			0607086840176050012745829940512669351340255776605063923326500235746768\
			2532374310004222447141820000150569200257907307135870590375790583196546\
			9128010576442044021361689686448073379253642027554432811384573029654738\
			3078418502700774711323764133172536234234574149578644484895450390248006\
			6116166289338522721532784374734848631663246408518328431964455453650689\
			8442519565407567055906470126647994864648150166847161605961719259553561\
			1285501711446715894282104789037944549954599610359279854765575016442679\
			2168230613666761669738753885256857659149289122839529868265651308460350\
			0816099242103629352076224118342410605210848333990826640871719526074806\
			7450645086068065068320145307706020784789967744941939873022138864987330\
			0402162317957167969085078599018397153640617466254108556042220461484604\
			2310700426302518115187134289420358604973593708206467003904135449176531\
			3359559410227516032325097975955361978124644871950768035111516999857683\
			7918122604511651102837922924304868423290529538094251509120612564950403\
			7134613099178202657501255043104658437610958948395571464625008843020168\
			0506404169322990593644244528311965667528452082996385707798419064872492\
			7923666626946599441064261345436946125975423314228545218713841932341884\
			9693185866802821968149546012739890871718413859466353743912692776142637\
			1243543007714400706075510696116058571352141114572746186416869642091089\
			4479165983028996997273582062057447681105276401712344902450651780129987\
			5917462521662603856939551722852396392723454848357146332104536193978434\
			1163281455871063299202962505697871614959319733648231024033562141411110\
			5479461461930120941554297719486376535320695108631020032277353988651779\
			2164015159051691558730221379654667964719180249873897742715169946708779\
			7187435291356005055609111604991764833222476269603313173554603845284047\
			9246633066574443144450009376285042939993927099409407588217795529379858\
			0380343335608232860339780023644074506694751176994053091637584704752141\
			9005830805251332053648182982378493594103878685067679569228256895373481\
			2379046575696178802930868781943944652118051204476651694583593043756365\
			9405242541869652504189897107019027155302991938136685327730655458779224\
			2642969798986409486882452930893379859750149249748539948066125475457714\
			9395111929283562102335201295434227863454776031155129017896565062464570\
			9300392395088189022789108528268627916541778448810037450839576198124902\
			0920634303581007557217546386326946635608535129098600580516476995946030\
			2053019766298695459320207417234455970779508265867103640609878989367805\
			7304428708248989988449759554844152647621330904746769195466281518740347\
			3923604480000000000000000000000000000000000000000000000000000000000000\
			0000000000000000000000000000000000000000000000000000000000000000000000\
			0000000000000000000000000000000000000000000000000000000000000000000000\
			0000000000000000000000000000000000000000000000000000000000000000000000\
			0000000000000000000000000000000000000000000000000000000000000000000000\
			0000000000000000000000000000000000000000000000000000000000000000000000\
			0000000000000000000000000000000000000000000000000000000000000000000000\
			0000000000000000000000000000000000000000000000000000000000000000000000\
			0000000000000000000000000000000000000000000000000000000000000000000000\
			0000000000000000000000000000000000000000000000000000000000000000000000\
			0000000000000000000000000000000000000000000000000000000000000000000000\
			0000000000000000000000000000000000000000000000000000000000000000000000\
			000000000000000000}		
	\end{minipage}
	\end{center}
\caption{\small A typical integer value arising in the linear recurrence of the color form factor $g_1C_F^3$.}
\label{fig:coeff}
\end{figure}
\end{center}

Besides the recurrences we also guessed linear differential equations for the form factors
\begin{equation}
f(s)=\sum_{n=0}^{\infty}F(n)s^n
\end{equation}
from the given coefficients $F(n)$ with $n=0,\dots,\rho$. As it turns out, to produce the linear differential equations for $f(s)$ required substantially more coefficients $F(n)$ than guessing the linear recurrences for $F(n)$. 
For instance, we needed around 37700 values to guess the linear differential equations of order $102$ for the color factor $C_F^3$  at the projection $g_1$ of the axial-vector form factor. Since we produced 40000 values directly from the large moment method for this particular instance, this calculation could be done without any extra complication. For all the other differential equations we used the available 25000 values (for simpler cases such as $\zeta_2$ and $\zeta_2$ around 10000 or 12000 values) to guess first the linear recurrences and to produce  afterwards 40000 values so that we were in the position to guess the differential equations.

Having obtained the governing differential equations, we executed Step~5 to evaluate the expansion coefficients of the form factors across the entire domain $s \in (-\infty, \infty)$. Special attention was given to critical physical points: the high-energy limit $s \to \pm\infty$, the threshold $s = 4$, and the pseudo-threshold $s = 16$. Leveraging the specialized techniques detailed in~\cite{BFMOS:26}, we successfully determined the expansion coefficients at $s = \pm\infty$ to an accuracy up to 4000 digits. Executing these computations efficiently posed a significant technical challenge due to the immense size of the underlying differential equations. High accuracy demanded dense grids of intermediate points for analytic continuation, as well as careful management of local poles that appear in individual differential equations but systematically cancel out in the final physical form factor. 
Remarkably, for the expansion around $s=\pm\infty$, this high-precision data allowed us to fit the coefficients via the PSLQ
algorithm~\cite{PSLQ1,PSLQ2,Bailey:1999nv} into linear combinations of multiple zeta values (MZVs) and three additional constants beyond the traditional MZV basis:
\begin{eqnarray}
	\label{eq:C1}
	{\tilde \kappa}_1 &=& \frac{3^5}{2^2} \HA_{\{3,0\},0,0}(1)
	\\
	{\tilde \kappa}_2 &=& \frac{3^5}{2^4} \HA_{\{3,0\},0,0}(1) \left[
	\frac{3}{2} \HA_{\{3,0\},0}(1) - \HA_{\{6,0\},0}(1)\right]
	\\
	\label{eq:C3}
	{\tilde \kappa}_3 &=& \frac{3^5}{2^2} \HA_{\{3,0\},0,0}(1) \HA_{\{\{4,-
		2,1\},0\},0,0}(1)
\end{eqnarray}
The first two constants are cyclotomic ones \cite{Ablinger:2011te}
and the third belongs to the class of quadratic form induced
polylogarithms
\cite{Ablinger:2021fnc}. We refer the reader to~\cite{BFMOS:26} for a comprehensive discussion on these constants and their symbolic derivation.

We remark that all these heavy calculations arising in Steps~1--5 have been carried by massive parallel processing both at the supercomputer MACH-2 (with 1728 processor cores and 20 TB of global shared memory) at JKU and the fast servers of DESY.

\section{Conclusion}

In Ref.~\cite{BFMOS:26}, we computed the gluonic contributions to three-loop heavy-quark form factors for vector, axial-vector, scalar, and pseudoscalar currents using advanced computer algebra techniques and high-performance automated guessing algorithms. In this note, we have summarized the core strategy of the large-moment method and highlighted its main computational challenges through concrete examples -- most notably involving one of the largest recurrences (and differential equations) ever guessed in contemporary particle physics problems. These high-precision evaluations deliver a fully analytic representation in the 
low-energy expansion around $s=0$ and the high energy expansion $s\to\pm\infty$ in terms of multiple zeta values (MZVs) and additional special constants, while enabling reliable numerical analytic continuations across the complete physical region $s \in (-\infty, \infty)$. Beyond their immediate application to precision QCD phenomenology, these results provide a crucial benchmark that demonstrates the power of combining cutting-edge differential equations, large-scale PSLQ integer-relation searches, and state-of-the-art symbolic summation methods.

\medskip

{\bf Acknowledgment.}
We thank DI J.~Messner from Zentraler Informationsdienst (ZID) Johannes
Kepler
University, Linz, for installing the parallelized guessing software
\cite{GSAGE}
at the supercomputer {\tt MACH-2} of JKU Linz and his help in
maintaining the runs
determining the largest differential and difference operators in this
project.
We thank M.~Kauers and K.~Sch\"onwald for discussions, and
D.~Broadhurst for
his interest in our work. This research was funded in whole or in part
by the
Austrian Science Fund (FWF) grants DOI 10.55776/P33530,
DOI 10.55776/PAT1332123, and DOI 10.55776/I6130.

\end{document}